\documentclass[12pt]{elsart}
\usepackage{amsmath,amssymb}
\usepackage{graphicx,psfrag,here,epsfig}
\usepackage{pstricks}
\usepackage{pst-node}
\usepackage{epsf}
\usepackage{citesort}
\allowdisplaybreaks[1]

\begin{document}
\newcommand{\co}{\; \; ,}
\def\words#1{\mbox{\small{\,#1}}}
\def\bea{\begin{eqnarray}}
\def\eea{\end{eqnarray}}
\def\eq{\begin{eqnarray}}
\def\en{\end{eqnarray}}
\def\be{\begin{equation}}
\def\ee{\end{equation}}
\newcommand{\ed}{\end{document}}

\newcommand{\nnnl}{\nonumber\\}
\newcommand{\fs}{\, . \,}
\def\query#1{\marginpar{\begin{flushleft}\footnotesize#1\end{flushleft}}}%

\runauthor{Colangelo, Gasser, Kubis and Rusetsky}
\renewcommand{\theequation}{\arabic{equation}}
\begin{frontmatter}

\begin{flushright}
HISKP-TH-06/09
\end{flushright}

\title{\Large\bf Cusps in \boldmath{$K\rightarrow 3\pi$} decays}

\author[Bern]{G.~Colangelo,}
\author[Bern]{J.~Gasser,}
\author[Bonn]{B.~Kubis,}
\author[Bonn]{A.~Rusetsky\thanksref{Tbilisi}}

\address[Bern]{Institute for Theoretical Physics, University of Bern,
Sidlerstr. 5, CH-3012 Bern, Switzerland}
\address[Bonn]{Helmholtz-Institut f\"ur Strahlen- und Kernphysik,
Universit\"at Bonn, Nussallee~14-16, D-53115 Bonn, Germany}

\thanks[Tbilisi]{On leave of absence from:
High Energy Physics Institute, Tbilisi State University,
University St.~9, 380086 Tbilisi, Georgia.}

\begin{abstract}
  The pion mass difference generates a pronounced cusp in
  $K\rightarrow3\pi$ decays. As has recently been pointed out by Cabibbo
  and Isidori, 
  an accurate measurement of the cusp may allow one to pin down the S-wave
  $\pi\pi$ scattering lengths to high precision. Here, we present and
  illustrate an effective field theory framework that allows one to determine
  the structure of this cusp in a straightforward manner. The
  strictures imposed by analyticity and unitarity are respected
automatically.
\end{abstract}


\end{frontmatter}

\noindent{\bf 1.}
 The S-wave $\pi\pi$ scattering lengths $a_0,a_2$ 
have been predicted with
 percent level accuracy some time ago
 \cite{Colangelo:2000jc,Colangelo:2001df},
 and first steps for an experimental verification of this prediction 
have been performed in Ref.~\cite{Pislak:2003sv,Adeva:2005pg}.
Recently, it has been pointed out by 
Cabibbo and Isidori \cite{Cabibbo:2004gq,Cabibbo:2005ez}
 that isospin violating effects
generate a pronounced cusp in $K\rightarrow 3\pi$ decays whose experimental
investigation may allow one to
 determine the combination $a_0-a_2$ with high precision. A
 first analysis of data based on this proposal 
has appeared \cite{Batley}.
(The strong impact of the unitarity cusp on $\pi^0\pi^0$ scattering
close to threshold was already mentioned in~\cite{MMS}.)
 In order for this program to be carried through successfully,
 one needs to determine
 the structure of the cusp with a precision that matches the experimental 
accuracy. In view of the large amount of data available 
\cite{Batley}, this
is a considerable task. In the present letter, we present a method which
-- we believe -- has the potential to achieve this goal.

In \cite{Cabibbo:2004gq,Cabibbo:2005ez}, the structure of the singularity
at the cusp is investigated using unitarity, analyticity and cluster
decomposition properties of the $S$-matrix.  In addition, an approximation
scheme is used, which consists in expanding the decay amplitude in powers
of $\pi\pi$ scattering lengths. The latest work \cite{Cabibbo:2005ez}
retains effects up to order (scattering lengths)$^2$ and omits explicit
electromagnetic effects. Here, we present a Lagrangian framework, which
automatically satisfies unitarity and analyticity constraints and, in
addition, allows one to include electromagnetic contributions in a standard
manner.  Specifically, we use a non-relativistic framework that has already
proven to be useful in the description of bound states
\cite{nrqfta,nrqft1,nrqft2,nrqft3,nrqft4,nrqft5,nrqft6,nrqft7,nrqft8,nrqft9,nrqft10,nrqft11,nrqft12,nrqft13,nrqft14,nrqft15,nrqftz}.
In this framework -- in contrast to relativistic field theory -- an
expansion in powers of scattering lengths emerges automatically from the
loop expansion. Moreover, it is a scheme that provides a proper power
counting.  [The low-energy expansion proposed here is closely related to
early work performed in the sixties by many authors (for a review see
\cite{anisovich}), who used $S$-matrix methods to investigate the
production of particles -- in parti\-cu\-lar also in $K\to3\pi$ decays --
in the threshold region.  The method presented here may be considered an
effective field theory realisation of these approaches.]

The strategy that we follow in this letter is the following.  First we
write down the most general non-relativistic Lagrangian relevant for this
decay and determine all
four-pion couplings therein through a matching procedure in terms of the
threshold parameters of $\pi\pi$ scattering.  In the next step, we evaluate
the $K\to 3\pi$ decay amplitudes to two loops. This results in an explicit
representation of the $S$-matrix elements which is valid in the whole decay
region and slightly beyond.  We propose to analyse the experimental data
with the use of this representation. We plan to include real and virtual
photon corrections at a later stage. On the other hand, we do keep the pion
and kaon masses at their physical values, which is a fully consistent
procedure in this framework.  This guarantees that the various branch
points and cusps occur at the proper place in the Mandelstam plane.
 
We display  the results without a detailed derivation, 
which will be provided in a forthcoming publication~\cite{big}.

\begin{sloppypar}
\noindent{\bf 2.}
We consider the neutral and charged decay modes
$K^+(p_K)\to\pi^0(p_1)\pi^0(p_2)\pi^+(p_3)$ and 
$K^+(p_K)\to\pi^+(p_1)\pi^+(p_2)\pi^-(p_3)$. The kinematical variables
are defined as usual: $s_i=(P_K-p_i)^2$ with $p_i^2=M_i^2\, ,~i=1,2,3$, where
$M_{\pi^+}\doteq M_\pi$ and $M_{\pi^0}$ denote the masses of the charged and
neutral pions, respectively, and $\Delta_\pi=M_\pi^2-M_{\pi^0}^2\neq 0$. 
In the centre-of-mass (CM) frame $P_K=(M_K,{\bf 0})$, with $M_K$ the
charged kaon mass,
\end{sloppypar}
\be\label{eq:kin1}
p_i^0=\frac{M_K^2+M_i^2-s_i}{2M_K}\, ,\quad
{\bf p}_i^2=\frac{\lambda(M_K^2,M_i^2,s_i)}{4M_K^2}\, ,
\ee
where $\lambda(x,y,z)=x^2+y^2+z^2-2xy-2xz-2yz$ is the triangle 
function. Below we also use the velocities $v_{jk}$ and kinetic energies $T_i$,
\be\label{eq:kin2}
v_{jk}^2(s_i)=\frac{\lambda(s_i,M_j^2,M_k^2)}{s_i^2}\, ,\quad
T_i=p_i^0-M_i\,\,.
\ee
\noindent {\bf 3.}
A non-relativistic approach to describe decays $K\to 3\pi$ can be 
justified, if the typical kinetic energies of the decay products 
are much smaller than the masses. This can be achieved by considering a world 
where the strange quark mass is taken to be smaller than its actual value.
Then, a consistent counting scheme arises, if one 
introduces a formal parameter $\epsilon$ and counts $T_i$ as a term 
of order $\epsilon^2$, the pion momenta as order $\epsilon$, 
whereas the pion and kaon masses are counted as $O(1)$.
From $\sum_iT_i=M_k-\sum_iM_i$, one concludes that the difference
$M_K-\sum_iM_i$ is then a quantity of order $\epsilon^2$ as well. 
The pion mass difference $\Delta_\pi$ is also counted as 
$O(\epsilon^2)$. The effective field theory 
framework, which we construct below, enables one
to obtain a systematic expansion of the amplitudes in $\epsilon$.
For sufficiently small $m_s$, the expansion in $\epsilon$ is expected to work
very well.

Together with $\epsilon$, our theory has another expansion parameter,
na\-me\-ly a characteristic size of the $\pi\pi$ threshold parameters,
which we denote generically as $a$.  In particular, the amplitudes in the
non-relativistic framework are given in form of an expansion in several
low-energy couplings $C_i,D_i$, which can be expressed in terms of the
threshold parameters of the relativistic $\pi\pi$ scattering amplitude.  We
expect the expansion in $a$ to converge rapidly because of the smallness of
the scattering lengths.  These two expansions are correlated \cite{big}:
because one-loop integrals are of order $\epsilon$, adding a pion loop
generated by a four-pion vertex increases both the order in $a$ and in
$\epsilon$ by one.  A consistent power counting is achieved: to a given
order in $a$ and in $\epsilon$, a well-defined finite number of diagrams
contribute.

Increasing now $m_s$ to its physical value again, 
convergence in the $\epsilon$-expansion
is not {\it a priori} evident, 
because $T_i/M_i$ can become as large as 0.4, and the 
corresponding maximal momentum ${\bf {|p|}}$ is then 
not much smaller than the pion mass.
 However, let us note the non-relativistic framework is only 
used to correctly reproduce
the non-analytic behaviour
of the decay amplitudes in the kinematical variables $s_1,s_2,s_3$, and to
thus provide a parametrisation consistent with unitarity and analyticity
 -- a trivial polynomial part in the amplitudes can be removed by a 
redefinition of the couplings in the Lagrangian.
In addition, from the analysis of the experimental data one  
knows~\cite{Cabibbo:2005ez} that in the
whole physical region the real part of the decay 
amplitude can be well approximated by a polynomial in  $s_1,s_2,s_3$ with
a maximum degree $2$. We interpret this fact as an experimental indication 
for a good convergence of the $\epsilon$-expansion for the 
quantities one is interested in.

We now proceed with the construction of the 
non-relativistic Lagrangian framework. In the decay amplitudes,
we shall restrict ourselves to terms up to and including 
 $O(\epsilon^2,a\epsilon^3,a^2\epsilon^2)$.

\noindent {\bf 4.}
It is convenient to formulate the non-relativistic approach in a manner
that describes the two-particle subsystems in a manifestly covariant
way~\cite{big}. We start with the $\pi\pi$ interaction and consider the
following five channels in $\pi^a\pi^b\to\pi^c\pi^d$: $(ab;cd)=$
(1)~$(00;00)$, (2)~$(+0;+0)$, (3)~$(+-;00)$, (4)~$(+-;+-)$, (5)~$(++;++)$.
The Lagrangian takes the form
\be\label{eq:l_pipi}
{\mathcal L}_{\pi\pi}=2\sum_\pm\Phi_\pm^\dagger W_\pm\bigl(i\partial_t-
W_\pm\bigr)\Phi_\pm 
+2\Phi_0^\dagger W_0\bigl(i\partial_t- W_0\bigr)\Phi_0
+\sum_{i=1}^5{\mathcal L}_{i}\, ,
\ee
where $\Phi_i$ is the non-relativistic pion field operator, $
W_\pm=\sqrt{M_\pi^2-\triangle}$, $ W_0=\sqrt{M_{\pi^0}^2-\triangle}$, with
$\triangle$ the Laplacian, and 
\eq\label{eq:l_1-5}
\hspace*{-15mm}{\mathcal L}_{i}&=&x_iC_i
\bigl(\Phi_c^\dagger\Phi_d^\dagger\Phi_a\Phi_b+h.c.\bigr)
+x_iD_i\Bigl\{\bigl( W_c\Phi_c^\dagger W_d\Phi_d^\dagger\Phi_a\Phi_b
+\Phi_c^\dagger\Phi_d^\dagger W_a\Phi_a W_b\Phi_b
\nonumber\\[1mm]
\hspace*{-1.mm}&+&\nabla\Phi_c^\dagger\nabla\Phi_d^\dagger\Phi_a\Phi_b
+\Phi_c^\dagger\Phi_d^\dagger\nabla\Phi_a\nabla\Phi_b
-h_i\Phi_c^\dagger\Phi_d^\dagger\Phi_a\Phi_b\bigr)+h.c.\Bigr\}+\ldots\,\,,
\en
with $h_1=2M_{\pi^0}^2$, $h_2=2M_\pi M_{\pi^0}$,
$h_3=3M_\pi^2-M_{\pi^0}^2$, $h_4=h_5=2M_\pi^2$.
The ellipsis stands for terms of order $\epsilon^4$ as well as for 
P-wave contributions, which occur at order 
$\epsilon^2$ in the $\pi\pi$ amplitude. They 
do not enter the $K\to 3\pi$ matrix elements
at the order of accuracy considered, so we omit them here.
The low-energy constants $C_i,D_i$ are matched to the physical scattering
lengths below.
To simplify the resulting expressions, we have furthermore 
introduced the scaling $x_1=x_5=1/4$, $x_2=x_3=x_4=1$.
 Finally, note that we omit local 6-pion couplings as well.
Their contribution to the $K\to 3\pi$ amplitude is purely imaginary in
the non-relativistic framework, and of order $\epsilon^4$, 
see also ~\cite{Cabibbo:2005ez,big}.

The pion propagator is given by 
\be\label{eq:propnew}
i\langle 0|T\Phi_a(x)\Phi_b^\dagger(y)|0\rangle= \delta_{ab}
\int\frac{d^4p}{(2\pi)^4}\,
\frac{{\rm e}^{-ip(x-y)}}{2w_a({\bf p})(w_a({\bf p})-p_0-i0)}\, ,\quad
a,b=\pm,0\, ,
\ee
and $w_\pm({\bf p})=\sqrt{M_\pi^2+{\bf p}^2}$,
$w_0({\bf p})=\sqrt{M_{\pi^0}^2+{\bf p}^2}$.
The loops are evaluated by using the following prescription:
one expands the square root in a series, 
calculates the emerging integrals using dimensional regularisation,
and sums up the series. In this manner 
one ensures that the power counting
 in the non-relativistic theory is not destroyed by the loop corrections.

All loops in $\pi\pi$ scattering can be expressed through the basic integral
\begin{equation}\label{eq:F}
\hspace*{-4.mm}J_{ab}(P^2)=\int\frac{d^Dl}{i(2\pi)^D}\,
\frac{1}{2w_a({\bf l})2w_b({\bf P}-{\bf l})}\,
\frac{1}{(w_a({\bf l})-l_0)(w_b({\bf P}-{\bf l})-P_0+l_0)}\, ,
\end{equation}
with $P^2=P_0^2-{\bf P}^2$. In the limit $D\to 4$,
\vspace*{-.1cm}
\begin{equation}\label{eq:functionJ}
J_{ab}(P^2)=\frac{i}{ 16\pi} v_{ab}(P^2) \,\, ,
\end{equation}

\vspace*{-.6cm}

which is a quantity of order $\epsilon$.

\noindent{\bf 5.}
The couplings $C_i,D_i$ can be expressed in terms of
 the threshold parameters of the underlying relativistic theory.
In the isospin symmetry 
limit, the expansion of the relativistic amplitude reads
\be
{\rm Re}\,\bar T_i(s,t)=
\bar A_i\Bigl\{1+\frac{\bar r_i}{4M_\pi^2}\,\bigl(s-4M_\pi^2\bigr)\Bigr\} 
+\ldots\,\,.
\ee 
The ellipsis stands for higher orders in $\epsilon$, 
as well as for P-wave
contributions. The bar indicates the isospin symmetric limit,
at $M_\pi=139.57$ MeV. In terms of the standard dimensionless scattering
lengths $a_0$ and $a_2$, one has
\bea\label{eq:isospin}
3\bar A_1&=&{N(a_0+2a_2)}{}\co\quad
2\bar A_2={Na_2}\co\quad
3\bar A_3={N(a_2-a_0)}\co\nnnl
6\bar A_4&=&{N(2a_0+a_2)}\co\quad
\bar A_5=Na_2 \; \; ; \qquad N=32 \pi\,\,,
\eea
with $a_0-a_2=0.265\pm0.004$ \cite{Colangelo:2000jc,Colangelo:2001df}.
The products $\bar A_i \bar r_i$ denote effective ranges.
Still in the isospin symmetry limit, the couplings $C_i,D_i$ are related to these
threshold parameters in the following manner,
\be\label{eq:pipimatching1}
 2\bar C_i=\bar A_i\,\,\,,\qquad 8 M_\pi^2 \bar D_i=\bar A_i\bar r_i\,\,.
\ee
Taking isospin breaking into account, one 
finds at leading order in chiral perturbation theory \cite{knechturech}
\be\label{eq:pipimatching2}
2C_{1,2,5}=\bar A_{1,2,5}(1-\eta),\,\,
2C_{3}=\bar A_3(1+\eta/3),\,\,
2C_{4}=\bar A_4(1+\eta),
\ee
where
$\eta=\Delta_\pi/M_{\pi}^2=6.5\times 10^{-2}$. Isospin breaking corrections in
$D_i$ do not contribute at the order considered here.

\begin{sloppypar}
\noindent{\bf 6.}
It remains to display the $K\rightarrow 3\pi$ 
Lagrangian,
\eq\label{eq:lag_K}
\hspace*{-4.mm}{\mathcal L}_K&=&2K^\dagger W_K\bigl(i\partial_t- W_K\bigr)K
 \nonumber \\[1mm]
\hspace*{-4.mm}&+&\frac{1}{2}\, G_0\,\bigl(K^\dagger\Phi_+\Phi_0^2+h.c.\bigr)
+\frac{1}{2}\, G_1\,\bigl(K^\dagger( W_+-M_\pi)\Phi_+\Phi_0^2+h.c.\bigr)
\\[1mm]
\hspace*{-4.mm}&+&\frac{1}{2}\, H_0\,\bigl(K^\dagger\Phi_-\Phi_+^2+h.c.\bigr)
+\frac{1}{2}\, H_1\,\bigl(K^\dagger( W_- -M_\pi)\Phi_-\Phi_+^2+h.c.\bigr)
+\ldots\, , \nonumber
\en
where $K$ denotes the non-relativistic field for the $K^+$ meson, 
$ W_K=\sqrt{M_K^2-\triangle}$, and the ellipsis stands for the
higher-order terms in $\epsilon$. Note that all couplings $G_i$, $H_i$ are
assumed to be real. Their contribution to the  decay matrix elements 
 at tree level is  provided below, in the amplitudes $A_{N,C}$.
\end{sloppypar}

The complete Lagrangian of the theory is
${\mathcal L}_K+{\mathcal L}_{\pi\pi}$.
 The tree-level expressions
 for the amplitudes, generated by ${\mathcal L}_K$, are
 modified by final state interactions of the pions, generated by  loops
 evaluated with ${\mathcal L}_{\pi\pi}$. We use the notation
\bea\label{eq:defexpand}
{\mathcal M}_{00+}={\mathcal M}_N^{\words{tree}}
                  +{\mathcal M}_N^{\words{1-loop}}
                  +{\mathcal M}_N^{\words{2-loops}}+\ldots\,\,\,\,
[K^+\to\pi^0\pi^0\pi^+]\,\,,\nonumber\\[1mm]
{\mathcal M}_{++-}={\mathcal M}_C^{\words{tree}}
                  +{\mathcal M}_C^{\words{1-loop}}
                  +{\mathcal M}_C^{\words{2-loops}}+\ldots\,\,\,\,
[K^+\to\pi^+\pi^+\pi^-]\,\,
\eea
for the decay amplitudes and the Condon-Shortley phase convention for 
the pions. Our amplitudes are
normalised such that the decay rates are given by
\be
d\Gamma=\frac{1}{2M_K}(2\pi)^4\delta^{(4)}(P_f-P_i)
{|\mathcal M|}^2\prod_{i=1}^3
\frac{d^3{\bf p}_i}{2(2\pi)^3p_i^0}\,.
\ee

\noindent{\bf 7.}
We now display the tree and one-loop results and modify the notation for
the couplings $C_i,D_i$ in order to make the formulae more transparent:
\be
(C_1,C_2,C_3,C_4,C_5)=(C_{00},C_{+0},C_{x},C_{+-},C_{++})\,,
\ee
and analogously for the $D_i$. We find
\eq\label{eq:1loop1}
{\mathcal M}_N^{\words{tree}} &=& A_N(s_3)\,\,,\nonumber\\[1mm]
 {\mathcal M}_N^{\words{1-loop}}
&=&B_{N1}(s_3)J_{+-}(s_3)+B_{N2}(s_3)J_{00}(s_3)
\nonumber\\[1mm]
&+&\{B_{N3}(s_1)J_{+0}(s_1) + (s_1\leftrightarrow s_2)\}\,\,,
\nonumber\\[1mm]
                  {\mathcal M}_C^{\words{tree}}&=& A_C(s_3)\,\,,
\nonumber\\[1mm]
 {\mathcal M}_C^{\words{1-loop}}
&=&
 B_{C1}(s_3)J_{++}(s_3)
\nonumber\\[1mm]
&+&\bigl\{B_{C2}(s_1)J_{+-}(s_1)
+B_{C3}(s_1)J_{00}(s_1)+(s_1\leftrightarrow s_2)\bigr\}\,\,,
\en
where\footnote{To render the formulae more compact, we keep some terms 
in Eqs.~(\ref{eq:1loop2}), (\ref{eq:1loop3}) that contribute  at 
order $a\epsilon^5$.}
\eq\label{eq:1loop2}
A_N(s_3)&=&
G_0+G_1\bigl(p_3^0-M_\pi\bigr)\, , \nonumber
\\[1mm]
B_{N1}(s_3)&=&2\bigl(C_x+D_x(s_3-\bar s_x)\bigr)
\biggl\{H_0+H_1\Bigl(\frac{p_1^0+p_2^0}{2}-M_\pi\Bigr)\biggr\}\, ,
\nonumber\\[1mm]
B_{N2}(s_3)&=&\bigl(C_{00}+D_{00}(s_3-\bar s_{00})\bigr)
\Bigl\{ G_0+G_1\bigl(p_3^0-M_\pi\bigr)\Bigr\}\, ,
\\[1mm]
B_{N3}(s_1)&=&2\bigl(C_{+0}+D_{+0}(s_1-\bar s_{+0})\bigr)
\biggl\{ \! G_0+G_1\Bigl(
\frac{p_2^0+p_3^0}{2}\Bigl(1+\frac{\Delta_\pi}{s_1}\Bigr)- \! M_\pi \!
\Bigr) \! \biggr\}\,,
\nonumber
\en
and
\eq\label{eq:1loop3}
A_C(s_3)&=&
H_0+H_1\bigl(p_3^0-M_\pi\bigr)\, , \nonumber
\\[1mm]
B_{C1}(s_3)&=&\bigl(C_{++}+D_{++}(s_3-\bar s_{++})\bigr)
\Bigl\{ H_0+H_1\bigl(p_3^0-M_\pi\bigr)\Bigr\}\, ,
\nonumber\\[1mm]
B_{C2}(s_1)&=&2\bigl(C_{+-}+D_{+-}(s_1-\bar s_{+-})\bigr)
\biggl\{ H_0+H_1\Bigl(\frac{p_2^0+p_3^0}{2}-M_\pi\Bigr)\biggr\}\, ,
\nonumber\\[1mm]
B_{C3}(s_1)&=&\bigl(C_x+D_x(s_1-\bar s_x)\bigr)
\Bigl\{ G_0+G_1\bigl(p_1^0-M_\pi\bigr)\Bigr\}\, .
\en
In the above expressions, $\bar s_i$ denotes the physical threshold in
the $i$ channel and $p_i^0$ are given by Eq.~(\ref{eq:kin1}).
Note that, according to this equation, the masses in the 
relation of $p_i^0$ to $s_i$ differ in the neutral and charged channels.

\begin{figure}[t]
\begin{center}
\includegraphics[width=9.cm]{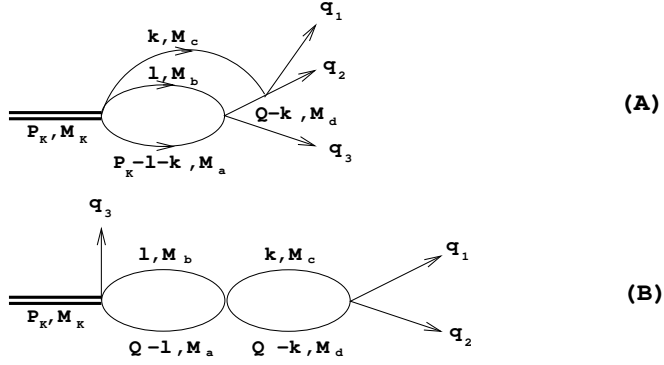}
\end{center}
\caption{Two topologically distinct non-relativistic two-loop graphs 
describing the final-state $\pi\pi$ rescattering in the decay $K\to 3\pi$, with
$Q^\mu=(q_1+q_2)^\mu$.}
\label{fig:2loop_text}
\end{figure}

\noindent {\bf 8.}
There are two topologically distinct two-loop graphs that describe
pion-pion rescattering in the final state, see Fig.~\ref{fig:2loop_text}.
 At the order of accuracy we are working, it is sufficient to consider the case
of non-derivative couplings. In this case, the contributions of 
both diagrams depend only on the variable $s$, where
\begin{equation}
Q^\mu=(q_1+q_2)^\mu \,\,,\,\,Q^2=s\,\, .
\end{equation}
The diagram
in  Fig.~\ref{fig:2loop_text}B, apart from a factor 
containing coupling constants, is given by a product of two one-loop
diagrams which were already calculated in Eq.~(\ref{eq:F}).
The non-trivial contribution from Fig.~\ref{fig:2loop_text}A
is proportional to
\eq\label{eq:twoloop_F}
{\mathcal M}(s)
&=&\int\frac{d^D l}{i(2\pi)^D}\,\frac{d^D k}{i(2\pi)^D}\,
\nonumber\\[1mm]
&\times&
\frac{1}{2w_a({\bf l}+{\bf k})}\,
\frac{1}{w_a({\bf l}+{\bf k})-M_K+l^0+k^0}\,
\frac{1}{2w_b({\bf l})}\, 
\frac{1}{w_b({\bf l})-l^0}\,
\nonumber\\[1mm]
&\times&\frac{1}{2w_c({\bf k})}\, 
\frac{1}{w_c({\bf k})-k^0}\,
\frac{1}{2w_d({\bf Q}-{\bf k})}\, 
\frac{1}{w_d({\bf Q}-{\bf k})-Q^0+k^0}\, .
\en
A detailed discussion of this integral will be provided in \cite{big}.
Here we simply note that
the most general
representation for diagram
Fig.~\ref{fig:2loop_text}A can be written in the form
\be\label{eq:mostgeneral}
{\mathcal M}(s)
=F\bigl(M_a,M_b,M_c,M_d;s\bigr)+P(s)J_{cd}(s)+P'(s)\, ,
\ee
where $F$ is ultraviolet-finite and
contains the full non-analytic behaviour of the two-loop diagram
in the low-energy domain. Further, $J_{cd}$ is the one-loop function
Eq.~(\ref{eq:functionJ}),
and $P,\; P'$ are real polynomials.
We have suppressed the dependence of 
 $F$ on the mass $M_i$ generated by  ${\bf Q}^2$, see below.

In the following, we use a simplified form of $F$, where 
part of its imaginary part is dropped -- this omission affects 
the decay width at order $a^3$ only and is therefore of no relevance here.
We use the integral representation \cite{big}
\eq\label{eq:functionF}
\hspace*{-.8cm}
&&F\bigl(M_a,M_b,M_c,M_d;s\bigr)=\nonumber\\[1mm]
&&\frac{{\mathcal N}_0}{64\pi^3\sqrt{s}}
\int_0^1\frac{dy}{\sqrt{y}}\,\frac{dg(y,s)}{dy}\,
\Bigl(\ln{g(y,s)}-\ln{g(y,\bar s)}\Bigr)+O(\epsilon^4)\, ,
\en
where
\eq
\hspace*{-.4cm}&&{\mathcal N}_0=\frac{M_K}{2\sqrt{s_0}}\,
\biggl(1-\frac{(M_a-M_b)^2}{s_0}\biggr)^{1/2}
\frac{1}{\bigl(2(M_K^2+M_c^2)-(M_a+M_b)^2-s_0\bigr)^{1/2}}\, ,
\nonumber\\[1mm]
\hspace*{-.4cm}&&s_0=M_K^2+M_c^2-2M_K\biggl( M_c^2+\frac{{\bf Q}^2(1+\delta)^2}{4}\biggr)^{1/2}\, ,
\nonumber\\[1mm]
\hspace*{-.4cm}&&g(y,s)=-(1-y)q_0^2-y\Delta^2+\frac{y(1-y){\bf Q}^2(1+\delta)^2}{4(1+y{\bf Q}^2/s)}-i0\, ,
\nonumber\\[1mm]
\hspace*{-.4cm}&&q_0^2=\frac{\lambda(s,M_c^2,M_d^2)}{4s}\, ,\quad\quad 
\bar s=(M_c+M_d)^2\, ,
\nonumber\\[1mm]
\hspace*{-.4cm}&&\Delta^2=\frac{\lambda(M_K^2,M_c^2,(M_a+M_b)^2)}{4M_K^2}\, ,\quad\quad
\delta=\frac{M_c^2-M_d^2}{s}\, .
\en
Approaching threshold from above, we find
\be
F\bigl(M_a,M_b,M_c,M_d;s\bigr)=-\frac{{q_0}}{128\pi^2(M_c+M_d)}
\frac{\lambda^{1/2}(\bar s_0,M_a^2,M_b^2)}{\bar s_0}+O(q_0^2)\, ,
\ee
where $\bar s_0$ denotes $s_0$ at $q_0^2=0$. 
This last relation shows that $F$ is of order $\epsilon^2$.

\begin{figure}[t]
\begin{center}
\includegraphics[width=13.5cm]{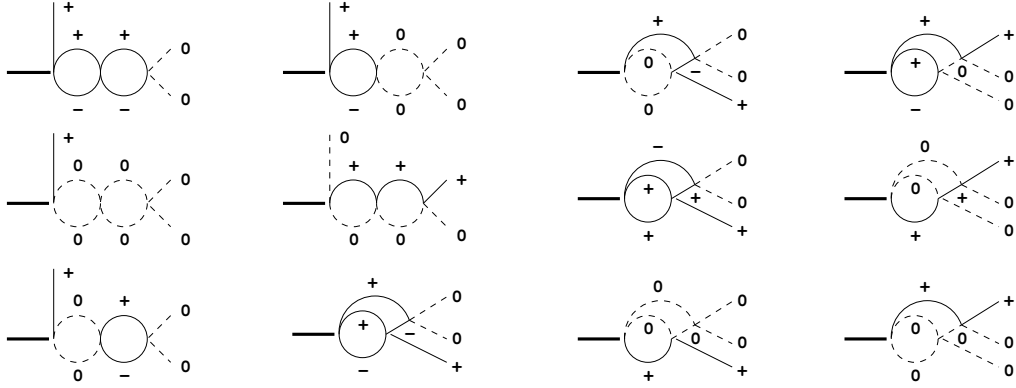}
\end{center}
\caption{Two-loop graphs contributing to the decay $K^+\to\pi^0\pi^0\pi^+$
in the non-relativistic effective theory. The graphs obtained by a 
permutation of identical particles in the final state are not
 shown.}
\label{fig:2loop_rep_n}
\end{figure}

\begin{figure}[t]
\begin{center}
\includegraphics[width=13.5cm]{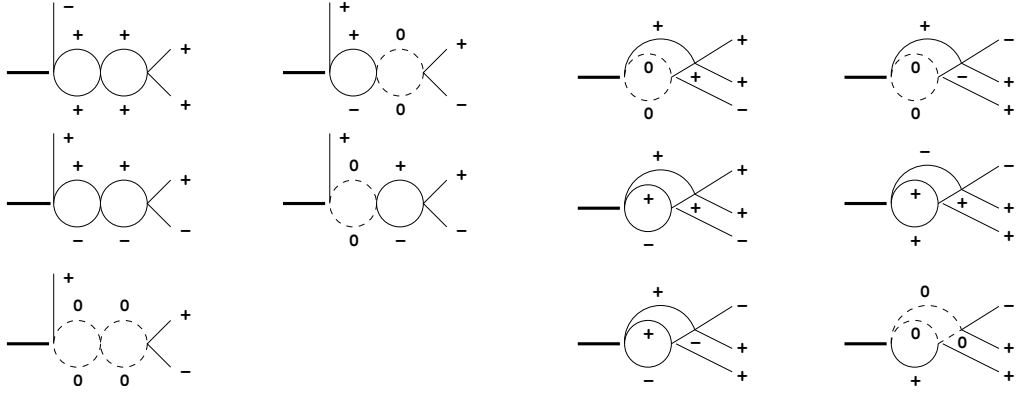}
\end{center}
\caption{Two-loop graphs contributing to the decay $K^+\to\pi^+\pi^+\pi^-$
in the non-relativistic effective theory. The graphs obtained by a
permutation of identical particles in the final state are not
 shown.}
\label{fig:2loop_rep_c}
\end{figure}

\noindent {\bf 9.}
Our prescription for the representation of the decay amplitudes at $O(a^2)$
is as follows: we evaluate the contributions from all the graphs displayed
in Fig.~\ref{fig:2loop_rep_n} and Fig.~\ref{fig:2loop_rep_c}.  Further, in
the graphs of the type 
Fig.~\ref{fig:2loop_text}A, we retain only the non-analytic piece $F$,
whereas the polynomials $P,\; P^\prime$ are included in the tree-level couplings
$G_i,H_i$. This choice of a particular representation of $F$ is equivalent
to a renormalisation prescription.

With this convention, we find for the amplitudes at order $a^2\epsilon^2$
\be\label{eq:rep}
{\mathcal M}_I^{\words{2-loops}}
={\mathcal M}_I^A(s_1,s_2,s_3)
+{\mathcal M}_I^B(s_1,s_2,s_3)\,\,;\,\, I=N,C\,\,,
\ee
where
\eq
\hspace*{-2.mm}{\mathcal M}_N^A
&=&4H_0C_{+-}C_xF_+\bigl(M_\pi,M_\pi,M_\pi,M_\pi;s_3\bigr)
\nonumber\\[1mm]
&+&2G_0C_x^2 F_+\bigl(M_{\pi^0},M_{\pi^0},M_\pi,M_\pi;s_3\bigr)
\nonumber\\[1mm]
&+&2H_0C_{++}C_xF_+\bigl(M_\pi,M_\pi,M_\pi,M_\pi;s_3\bigr)
\nonumber\\[1mm]
&+&4G_0C_{00}C_{+0}F_+\bigl(M_\pi,M_{\pi^0},M_{\pi^0},M_{\pi^0};s_3\bigr)
\nonumber\\[1mm]
&+&\Bigl\{4H_0C_xC_{+0}F_0\bigl(M_\pi,M_\pi,M_\pi,M_{\pi^0};s_1\bigr)
\nonumber\\[1mm]
&+&4G_0C_{+0}^2F_0\bigl(M_{\pi^0},M_\pi,M_{\pi^0},M_\pi;s_1\bigr)
\nonumber\\[1mm]
&+&2G_0C_{00}C_{+0}F_0\bigl(M_{\pi^0},M_{\pi^0},M_\pi,M_{\pi^0};s_1\bigr)
+(s_1\leftrightarrow s_2)\Bigr\}\,\,, \nonumber \\
&& \nonumber \\
\label{eq:0B}
{\mathcal M}_N^B
&=&4H_0C_xC_{+-}J^2_{+-}(s_3)
+G_0C_{00}^2J^2_{00}(s_3)
\nonumber\\[1mm]
&+&2\bigl[G_0C_x^2+H_0C_xC_{00}\bigr]J_{+-}(s_3)J_{00}(s_3)
\nonumber\\[1mm]
&+&\Bigl\{4G_0C_{+0}^2J^{\,\,2}_{+0}(s_1)
+(s_1\leftrightarrow s_2)\Bigr\}\, ,
\en
\eq
\hspace*{-5.mm}{\mathcal M}_C^A
&=&2G_0C_xC_{++} F_-\bigl(M_{\pi^0},M_{\pi^0},M_\pi,M_\pi;s_3\bigr)\, ,
\nonumber\\[1mm]
&+&4H_0C_{+-}C_{++} F_-\bigl(M_\pi,M_\pi,M_\pi,M_\pi;s_3\bigr)
\nonumber\\[1mm]
&+&\Bigl\{4H_0C_{+-}^2 F_+\bigl(M_\pi,M_\pi,M_\pi,M_\pi;s_1\bigr)
\nonumber\\[1mm]
&+&2G_0C_xC_{+-} F_+\bigl(M_{\pi^0},M_{\pi^0},M_\pi,M_\pi;s_1\bigr)
\nonumber\\[1mm]
&+&2H_0C_{++}C_{+-} F_+\bigl(M_\pi,M_\pi,M_\pi,M_\pi;s_1\bigr)
\nonumber\\[1mm]
&+&4G_0C_{+0} C_x F_+\bigl(M_\pi,M_{\pi^0},M_{\pi^0},M_{\pi^0};s_1\bigr)
+ (s_1\leftrightarrow s_2)\Bigr\}\co \nonumber \\
&& \nonumber \\
\label{eq:+B}
{\mathcal M}_C^B
&=&H_0C_{++}^2J^2_{++}(s_3)\nonumber\\[1mm]
&+&\Bigl\{ 4H_0C_{+-}^2J^2_{+-}(s_1)+G_0C_xC_{00}
J^2_{00}(s_1)
\nonumber\\[1mm]
&+&2\bigl[H_0C_x^2+G_0C_x C_{+-}\bigr]J_{+-}(s_1)J_{00}(s_1)
+(s_1\leftrightarrow s_2)\Bigr\}\,\,.
\en
\begin{sloppypar}
Here, $F_i(\ldots;s)$ stands for the integral $F(\ldots;s)$, 
evaluated at ${\bf
    Q}^2=\lambda(M_K^2,M_{\pi^i}^2,s)/4M_K^2$, with $i=\pm,0$.
\end{sloppypar}

\noindent{\bf 10.}
The decay amplitudes depend on the four {\em real} $K\to 3\pi$ coupling
constants $H_i,G_i$ and on the threshold parameters for $\pi\pi$
scattering. Combining the tree and one-loop result
Eqs.~(\ref{eq:1loop1})--(\ref{eq:1loop3}) with the two-loop contributions
Eqs.~(\ref{eq:rep})--(\ref{eq:+B}), we obtain the neutral and charged decay
amplitudes up to and 
including terms of order $\epsilon^2,a\epsilon^3$ and $a^2\epsilon^2$,
expressed in terms of the one- and two-loop integrals $J$ and $F$ displayed
in Eqs.~(\ref{eq:functionJ}) and (\ref{eq:functionF}), respectively.  This
representation is valid in the whole decay region, and is the main result
of this article.

The decay amplitude $K^+\to\pi^0\pi^0\pi^+$ obeys what we refer to as the
{\it threshold theorem}: the coefficient of the leading non-analytic piece,
which is proportional to $v_{+-}(s_3)$, is given by a product of two
factors, the decay amplitude $K^+\to\pi^+\pi^+\pi^-$ and the scattering
amplitude $\pi^+\pi^-\to\pi^0\pi^0$, both evaluated at threshold.  Thus,
the heuristic argumentation of Ref.~\cite{Cabibbo:2004gq}, which serves as
a cornerstone of the whole method, is confirmed in the effective field
theory framework. This threshold theorem has its analogue in hadronic
atoms, viz., in the modification of energy levels and decay widths through
hadronic interactions
\cite{nrqft6,nrqft7,nrqft8,nrqft9,nrqft10,nrqft11,nrqft12,nrqft13,nrqft14,nrqft15,nrqftz}.
Of course, aside from the determination of the leading term in $v_{+-}$,
our approach also allows a systematic evaluation of higher-order
contributions $v_{+-}^3,v_{+-}^5\ldots$\,.

\noindent {\bf 11.}
We now compare the content of this letter with the recent work of Cabibbo
and Isidori~\cite{Cabibbo:2005ez} (CI), who have proposed an alternative
representation for the $K\to 3\pi$ decay amplitudes. 
Assuming certain analytic properties of the decay amplitudes, CI derive a
representation of the amplitudes up to and including terms of order $a^2$,
using analyticity, unitarity, and cluster decomposition properties of the
$S$ matrix.

Conceptual aspects of our methods have already been compared in the
introduction. Here we add that, firstly, we do provide an explicit
representation of the decay amplitudes that is valid in the whole decay
region, including all powers of velocities generated by the graphs
considered. Secondly, we note that our explicit two-loop calculation
confirms the expected \cite{Cabibbo:2005ez} analytic properties of the
amplitudes in the vicinity of the $\pi^+\pi^-$ threshold. On the other
hand, away from this threshold, but still in the physical decay region,
branch points develop \cite{big}, contrary to the expectations spelled out
in \cite{Cabibbo:2005ez}.

After these general remarks we compare the amplitudes in more detail at one
and two loops. In the actual calculations in Ref.~\cite{Cabibbo:2005ez},
an approximation has been used: the angular integrals have been replaced by
averages, where the integrand is evaluated at a certain value of
$\cos\theta$. As CI note, this approximation is exact, if the integrand is
at most a linear function in $\cos\theta$. Since this is true at
$O(a\epsilon^2)$, one expects that our results at this order algebraically
agree with CI. We have checked that this is indeed the case (up to a few
typos \cite{isidoriprivate}).

At two loops, our results are algebraically different from those of
Ref.~\cite{Cabibbo:2005ez}. The reason can be traced back to the following.
At two loops, the discontinuity cannot, in general, be obtained from an
integration over $\cos\theta$ without further ado: the path of integration
has to be deformed into the complex plane -- it does not simply run from
$\cos\theta=-1$ to $\cos\theta=1$ along the real axis \cite{anisovich}. In
fact, near the pseudothreshold $s_3=(M_K-M_\pi)^2$, the deformed path runs
to infinity, thus generating an infinity in the discontinuity there. In the
case where all internal masses are equal to the charged pion mass,
integrating $\cos\theta$ along the real axis does generate the correct
discontinuity up to $s_3=(M_K^2-M_\pi^2)/2$. In the case where two neutral
pions are running in the inner loop, the situation is more complicated,
because an anomalous threshold develops in the lower half plane. Still,
integrating along the path mentioned generates the correct discontinuity
near threshold.  [These difficulties do not arise in our approach
-- the function $F$ in Eq.~(\ref{eq:functionF}) is smooth on the upper rim
of the cut, while its discontinuity develops the singularities mentioned,
at the positions predicted by the Landau equations
\cite{landau,smatrix,big}.]

A replacement of the angular integrals by an average
\cite{Cabibbo:2005ez}, where the integrand is evaluated at a certain value
of $\cos\theta$, can therefore be a reasonable approximation only in the
vicinity of the threshold $s_3=4M_\pi^2$. Indeed, our result agrees (up to
a typo~\cite{isidoriprivate}) with Ref.~\cite{Cabibbo:2005ez} to
the order considered here, at threshold. Away from threshold, the
expressions differ.

Finally, we shortly comment on the (revised) article by Gamiz
et al. \cite{gamizetal}, which appeared only very recently. It is the aim
of that article to provide an error analysis of the procedure proposed in
\cite{Cabibbo:2004gq,Cabibbo:2005ez}. The authors investigate the process
$K\rightarrow 3\pi$ in the framework of chiral perturbation theory, and
approximate two-loop graphs by retaining their discontinuity only -- an
approach which generates fake singularities in the transition amplitude, as
just mentioned. An ad hoc prescription is invoked in \cite{gamizetal} to
avoid these singularities when investigating the cusp, see e.g. their
comment after Eq.~(4.20).

\noindent {\bf 12.}
In summary, we have investigated $K\to 3\pi$ decays within a
non-relativistic effective Lagrangian framework.  The amplitudes are
calculated in a systematic double expansion in the kinetic energies of the
decay products (which we count as terms of order $\epsilon^2$), and in the
threshold parameters of elastic $\pi\pi$ scattering (which are generically
denoted by $a$).  We provide an explicit representation of the amplitudes
at order $\epsilon^2, a\epsilon^3, a^2\epsilon^2$ -- valid in the whole
decay region -- in terms of the (real) $K\to 3\pi$ coupling constants
$G_i,H_i$ and of the threshold parameters $a$.

Our amplitudes agree with the ones of Cabibbo and
Isidori \cite{Cabibbo:2005ez}, up to and including terms of 
order $a\epsilon^3$.
On the other hand, at order $a^2$, they differ
away from threshold. We propose to repeat the
analysis of the experimental data of the NA48/2 collaboration
\cite{Batley} with our representation of the amplitudes, for
several reasons: i) In view of the aimed precision, one ought to examine
the importance of the mentioned differences in the determination of
$\pi\pi$ scattering lengths. ii) It would be useful to extend the fit to
the full decay region, and to the charged decay modes $K^+\rightarrow
\pi^+\pi^+\pi^-$ as well, in order to determine a maximal set of $\pi\pi$
threshold parameters. iii) As was already pointed out in
\cite{Cabibbo:2005ez}, cusps also occur at the border of the Dalitz plot.
Investigating data in those regions may allow one to determine different
combinations of scattering lengths.

It remains to investigate the importance of higher orders in the low-energy
expansion, and to apply radiative corrections, which can be evaluated
in the field-theoretical framework used here in a standard manner. 
The effects generated by the $\pi^+\pi^-$ bound state at the $\pi^+\pi^-$
threshold can also be investigated within
the same approach (see, e.g., 
\cite{nrqft6,nrqft7,nrqft8,nrqft9,nrqft10,nrqft11,nrqft12,nrqft13,nrqft14}).
We plan to include these effects in forthcoming publications.

\begin{sloppypar}
{\it Acknowledgements.}
It is a pleasure
to thank N.~Cabibbo, J.~Donoghue, G.~Isidori and J.~Prades for enjoyable discussions.
Partial financial support under the EU Integrated Infrastructure
Initiative Hadron Physics Project (contract number RII3-CT-2004-506078)
and DFG (SFB/TR 16, ``Subnuclear Structure of Matter'') is gratefully
acknowledged. This work was  supported  by the Swiss
National Science Foundation, by RTN, BBW-Contract No. 01.0357,
and EC-Contract  HPRN--CT2002--00311 (EURIDICE).
 One of us (J.G.) is grateful to the Alexander von Humboldt-Stiftung and to 
the Helmholtz-Gemeinschaft for the award of  a prize 
 that allowed him to stay at the HISKP at the University of Bonn, 
where part of this work was performed. 
He also thanks the HISKP for the warm hospitality during these stays.
\end{sloppypar}

\ed